\input{aipcheck}

\documentclass[
    ,final            
  ]
  {aipproc}

\layoutstyle{8x11single}

\begin{document}

\title{Significance of neutrino cross-sections for astrophysics}


\author{A.B. Balantekin}{
  address={Department of Physics,
University of Wisconsin, Madison, WI 53706, USA}
}

\begin{abstract}
The heavens are full of neutrinos, but so far we have observed a small subset of those. Performing such observations and analyzing the data from them require a thorough understanding of neutrino interactions. In this short review, the current theoretical status of neutrino cross sections are briefly summarized and novel aspects of neutrino interactions that arise in cosmic sites are discussed. 
\end{abstract}

\maketitle


\section{Neutrino Cross Sections}

Our first observation of the neutrino sky did not take place until the efforts of Davis and his collaborators in the Homestake mine in the late 1960's. Since then, we have seen neutrinos from two (and only two) heavenly objects: a main-sequence star (the Sun) and a core-collapse supernova (SN1987A). 
These low-energy astrophysical neutrinos provide us tools for a new kind of astronomy looking at the interior of compact objects. They help us explore fascinating phenomena in the Cosmos as diverse as the birth of new stars and the origin of elements. Thus neutrino physics provides a new tool to look at the cosmos complementary to other tools already in place such as electromagnetic telescopes and gravitational wave detectors such as LIGO.  

To fully assess the role of neutrinos in astrophysical phenomena we need to have a good handle on neutrino interactions. In most cases of astrophysical interest it is not possible to directly measure the needed cross-sections. A fundamental approach would be to use effective field theories:  For low-energy neutrino-deuteron scattering (below the pion threshold),  $^3S_1 \rightarrow ^1S_0$ transition dominates and one only needs the coefficient of the two-body counter term, so called L$_{1A}$, specifying isovector two-body axial current \cite{Butler:1999sv}. 
The term L$_{1A}$ can be obtained either by comparing the cross section $\sigma(E) = \sigma_0(E) + L_{1A} \sigma_1(E)$ with cross-section calculated using other approaches or measured experimentally 
(e.g. use solar neutrinos as a source) \cite{Butler:2002cw}. Difficulties with implementing three-body forces practically restrict the use of effective field theory approach only to reactions involving the deuteron. For scattering on nuclei beyond deuteron, other approaches are needed.  
At the lowest energies, relevant to the Sun and supernovae,  theoretical tools are Shell Model (SM) and Random Phase Approximation (RPA) or quasiparticle RPA (QRPA). At the lowest energies Shell Model is the best approach. As the incoming neutrino energy increases, the contribution of the states which are not well-known increase, including first- and even second-forbidden transitions \cite{Lazauskas:2007bs}. At higher energies where the rate is sensitive to total strength and the energy of giant resonances there is a tendency to use RPA.  Both techniques start with the mean-field approximation for nucleons in nuclei. However, during the past decades, electron scattering experiments have provided overwhelming evidence of correlations in nuclei, requiring the use of realistic nucleon-nucleon potentials \cite{Benhar:2006wy}. Indeed, importance of short-rage correlations in QRPA was emphasized long time ago \cite{Engel:1988au}. A study of neutrino scattering off nuclei at intermediate or higher energies need to include the effects of those correlations. It should be noted that in astrophysical settings new final-state effects may come into play; for example, in a core-collapse supernova neutrino capture reactions may be influenced by the Pauli-blocking by other electrons present \cite{Minato:2006zk}. 

To experimentally study neutrino cross-sections at astrophysically relevant low-energies 
(i.e. to measure the spin-isospin response of the nuclei to weak probes)  one clearly needs neutrinos with  varying energies. With low-energy beta-beams it would be possible to do a  systematic study of spin-isospin response of nuclei \cite{Volpe:2006in}. Such a low-energy beta-beam facility could also enable to test other aspects of neutrino interactions. For example, one can directly measure the weak-magnetism contribution to the cross-section for the inverse beta decay, providing another test of the conserved vector current hypothesis \cite{Balantekin:2006ga}. (Weak magnetism contributions may, for example, increase the antineutrino mean free path in a supernova \cite{Horowitz:2001xf}). Other tests of electroweak physics \cite{Balantekin:2005md} may also be feasible\footnote{Note that  it is also possible to utilize low-energy neutrinos at off-axis from  standard beta-beams to do some of these experiments \cite{Lazauskas:2007va}. }.  
In addition, low-energy beta beams may provide a suitable venue to mimic the neutrino flux from a core-collapse supernova \cite{Jachowicz:2006xx}.  It should also be noted that beta-beams and other future neutrino facilities would help us to probe new physics contributions to the neutrino interactions at higher energies \cite{Balantekin:2008rc,Balantekin:2008vw}. 

\section{Solar Neutrinos}

The field of neutrino astrophysics started in the 1960's with an attempt to measure the solar neutrino flux in the Homestake mine. After three decades of dedicated experimental efforts we now have precise measurements of not only the total solar neutrino flux, but also of its individual flavor components. This was hailed as a crucial test of the theory of main sequence stellar evolution in addition to being a measurement of neutrino mass differences and mixings. In parallel to the developments in solar neutrino physics, careful helioseismic observations of the solar acoustic modes were able to map the sound-speed profile within the Sun. 

Abundances of elements such as C and N in the Sun are best inferred from the photospheric absorption lines. Until recently, input from such observations \cite{Anders:1989zg,Grevesse:1998bj} into the Standard Solar Model yielded a sound speed profile in good agreement with helioseismic observations. 
However, more recent observations suggest a significant decrease in solar metalicity \cite{Asplund:2003ws}: the Sun is no longer an {\it odd} star enriched in heavy elements. The resulting reduction in the core temperature lowers the component of the solar neutrino flux most sensitive to core temperature, the $^8$B flux \cite{TurckChieze:2004mg}. However, since the CN cycle neutrinos have a stronger dependence on core temperature, a better measurement of their flux would lead to a better understanding of the solar metalicity \cite{Haxton:2008yv}.  A more precise measurement of the CNO contribution will not only provide a test of energy generation during the main sequence of stellar evolution, but also shed light on the solar metalicity. Present uncertainty is very big, but a few percent of accuracy is within reach. An accurate measurement of total solar neutrino luminosity will also help us to probe if the nuclear fusion reactions are the only source of solar energy.  

Related open questions include searches for subdominant neutrino sources (testing the relation between solar photon and neutrino luminosities), neutrino magnetic moments and their effects on the solar neutrino flux, using neutrinos to measure solar properties such as the density scale height and to search for physics beyond both the Standard Model of the Sun and the Standard Model of particle physics. 
All these questions require an accurate knowledge of either production or detection cross sections for neutrinos. 

\section{Neutrinos from Core-Collapse Supernovae}

A core-collapse supernova event can be considered primarily a neutrino affair \cite{Balantekin:2003ip}.  Almost all of the gravitational binding energy of the progenitor star and about 10\% of the rest mass is radiated as neutrinos of all three flavors. Although neutrinos from a core-collapse supernova were already observed, it would be very useful to see a second one besides SN1987A. In addition, ability to observe supernovae in nearby galaxies would greatly improve the statistics.  Observation of all the past supernovae through their neutrino signatures (the diffuse supernova neutrino flux) gives us a handle on the star-formation rate. 

Supernovae are candidate sites for r-process nucleosynthesis,  which requires a neutron-rich environment, i.e. the ratio of electrons to baryons, $Y_e$, should be well below $0.5$. Electron neutrinos energetic enough to push the baryons with the neutrino-driven wind also tend to convert neutrons to protons. Because of the large alpha-particle binding energy, those protons will capture additional neutrons to form alpha particles, pushing $Y_e$ towards $0.5$ \cite{Meyer:1998sn}. 
Active-to-sterile matter-enhanced neutrino oscillations  \cite{Fetter:2002xx} and active-to-sterile magnetic moment conversion  \cite{Balantekin:2007xq} are proposed as possible ways to get around this "alpha effect", although the latter scenario likely requires too large a neutrino magnetic moment. 
 
For the matter-enhanced neutrino oscillations (MSW effect) the matter potential neutrinos experience is provided by the coherent forward scattering of neutrinos off the electrons (or positrons) in dense 
matter\footnote{There is a similar term with Z-exchange. But since it is the same for all neutrino flavors, it does not contribute to phase differences at the tree-level except when one invokes sterile neutrinos.}. 
If the neutrino density itself is also very high then one has to consider the effects of neutrinos scattering off other neutrinos. This is the case for a core-collapse supernova.  Yields of r-process nucleosynthesis are determined by neutron-to-proton ratio, or equivalently by the electron fraction. Interactions of the neutrinos and antineutrinos streaming out of the core both with nucleons and seed nuclei determine this ratio. Hence it is crucial to understand (and measure) neutrino-nucleon as well as neutrino-nucleus cross-sections. Furthermore, before these neutrinos reach the r-process region they undergo matter-enhanced neutrino oscillations as well as coherently scatter over other neutrinos.  Many-body behavior of this neutrino gas is not completely explored, but may have significant impact on r-process nucleosynthesis.

For the illustrative case of two flavors of neutrinos (and no antineutrinos) one can define the SU(2) operators
\begin{equation}\label{1}
J_+(p)= a_x^\dagger(p) a_e(p), \ \ \ \
J_-(p)=a_e^\dagger(p) a_x(p), \ \ \ \
J_0(p)=\frac{1}{2}\left(a_x^\dagger(p)a_x(p)-a_e^\dagger(p)a_e(p)
\right) . 
\end{equation}
Forward scattering of neutrinos from other neutrinos is described by the Hamiltonian \cite{Balantekin:2006tg} 
\begin{equation}\label{2}
H_{\nu \nu} = \frac{\sqrt{2} G_F}{V} \int d^3p \> d^3q \>  R_{pq}
\> {\bf J}(p) \cdot {\bf J}(q) .
\end{equation}
where we defined
\begin{equation}
R_{pq} = (1-\cos\vartheta_{pq}) . \label{3} 
\end{equation} 
It is possible to include all three flavors and antineutrinos in these equations. 
Eq. (\ref{2}), together with the MSW Hamiltonian define a complicated many-body problem, the exact solution of which is still lacking. However, its saddle-point solution \cite{Balantekin:2006tg}, where the neutrinos interact with a mean field, is commonly used to study neutrino propagation with
neutrino-neutrino interactions (see Refs. \cite{Qian:1994wh}
through \cite{Kneller:2009vd}). Numerical studies indicate that the addition of a neutrino
background results in very interesting physical effects such as
coherent flavor transformation. This discussion also illustrates that astrophysical extremes allow testing neutrino properties in ways that cannot be done elsewhere. 

Calculations with neutrino-neutrino interactions find large-scale, collective flavor oscillations deep in the supernova envelope, which is sensitive to the value of $\theta_{13}$. Such oscillations may also effect r-process nucleosynthesis. As an example, results from a neutrino mean-field calculation of electron fraction is shown in Figure 1 \cite{Balantekin:2004ug}.


\begin{figure}
  \includegraphics[height=.3\textheight]{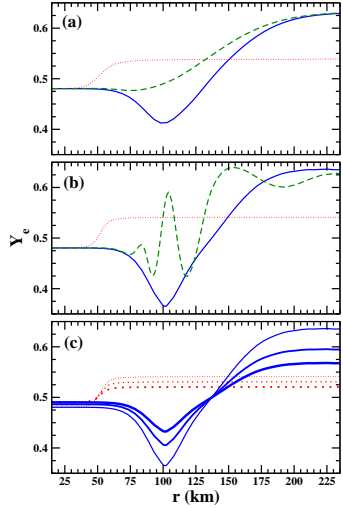}
  \caption{Electron fraction as a function of the distance from the core in a core-collapse supernova with neutrino-neutrino interactions included (from Ref. \cite{Balantekin:2004ug}). $\delta m^2_{13} \sim 3 \times 10^{-3} {\rm eV}^2$, luminosities are $L^{51} = 0.001, 0.1 \>{\rm and}\> 100$ (solid, dashed and dotted lines). a) $\theta_{13} \sim \pi/10$, b) $\theta_{13} \sim \pi/20$, c) same as b), but with the alpha effect when the alpha fraction is 0, 0.3, and 0.5 (thin, medium, and thick lines).}
\end{figure}

\section{Conclusions}

Neutrino cross sections are often a crucial input in understanding a broad range of phenomena ranging from stellar evolution to core-collapse supernovae. 
Understanding neutrino interactions from very low to very high energies are often crucial in interpreting the data from a variety of astrophysical sources. 
 At low energies (supernova and tail of atmospheric neutrinos) calculational tools are still being developed. At very high energies physics beyond the Standard Model is very likely to contribute. 
 At astrophysical sites novel forms of neutrino interactions emerge such as collective behavior driven by neutrino-neutrino interactions. 


\begin{theacknowledgments}
This work was supported in part
by the U.S. National Science Foundation Grants No.\ PHY-0555231 
and PHY-0855082 
and
in part by the University of Wisconsin Research Committee with funds
granted by the Wisconsin Alumni Research Foundation.
\end{theacknowledgments}

\end{document}